\documentclass[12pt]{article}

\pdfoutput=1

\usepackage{cite}

\usepackage{amssymb,latexsym}

\usepackage{epstopdf}

\usepackage{graphics,epsfig}

\usepackage{color}

\let\a=\alpha \let\b=\beta \let\g=\gamma \let\d=\delta \let\e=\epsilon
\let\z=\zeta  \let\th=\theta  \let\k=\kappa
\let\l=\lambda \let\m=\mu \let\n=\nu \let\x=\xi \let\p=\pi 
\let\s=\sigma   \let\f=\phi  
 
        \let\Th=\Theta \let\L=\Lambda
\let\X=\Xi  \let\S=\Sigma  \let\Y=\Psi
 
\let\la=\label  
  
\def\nn{\nonumber} \def\bd{\begin{document}} \def\ed{\end{document}}
\def\ds{\documentstyle} \let\fr=\frac \let\bl=\bigl \let\br=\bigr
\let\Br=\Bigr \let\Bl=\Bigl
\let\bm=\bibitem
\let\na=\nabla
\def\tU{{\widetilde U}}
\let\pa=\partial \let\ov=\overline
\def\ie{{\it i.e.\ }}
\newcommand{\be}{\begin{equation}}
\newcommand{\ee}{\end{equation}}
\def\ba{\begin{array}}
\def\ea{\end{array}}
\def\ft#1#2{{\textstyle{{\scriptstyle #1}\over {\scriptstyle #2}}}}
\def\fft#1#2{{#1 \over #2}}
\def\F#1#2{{ F_{#1}^{(#2)} }}
\def\cF#1#2{{ {\cal F}_{#1}^{(#2)} }}

\def\R{{\bf R}}
\def\sst#1{{\scriptscriptstyle #1}}
\def\oneone{\rlap 1\mkern4mu{\rm l}}
\def\e7{E_{7(+7)}}
\def\td{\tilde}
\def\wtd{\widetilde}
\def\im{{\rm i}}
\def\bog{Bogomol'nyi\ }
\newcommand{\ho}[1]{$\, ^{#1}$}
\newcommand{\hoch}[1]{$\, ^{#1}$}
\newcommand{\bea}{\begin{eqnarray}}
\newcommand{\eea}{\end{eqnarray}}
\newcommand{\ra}{\rightarrow}
\newcommand{\lra}{\longrightarrow}
\newcommand{\Lra}{\Leftrightarrow}
\newcommand{\ap}{\alpha^\prime}
\newcommand{\bp}{\tilde \beta^\prime}
\newcommand{\cB}{{\cal B}}
\newcommand{\cO}{{\cal O}}
\newcommand{\vecx}{\vec{x}}
\newcommand{\vecy}{\vec{y}}
\newcommand{\vecp}{\vec{p}}
\newcommand{\vecq}{\vec{q}}
\newcommand{\tr}{{\rm tr} }
\newcommand{\Tr}{{\rm Tr} }
\newcommand{\NP}{Nucl. Phys. }

\newcommand{\cL}{{\cal L}}
\newcommand{\cA}{{\cal A}}
\newcommand{\cT}{{\cal T}}
\newcommand{\cD}{{\cal D}}
\newcommand{\cH}{{\cal H}}

\def\sst#1{{\scriptscriptstyle #1}}
\def\0{{\sst{(0)}}}
\def\1{{\sst{(1)}}}
\def\2{{\sst{(2)}}}
\def\3{{\sst{(3)}}}
\def\4{{\sst{(4)}}}
\def\5{{\sst{(5)}}}
\def\6{{\sst{(6)}}}
\def\7{{\sst{(7)}}}
\def\8{{\sst{(8)}}}
\def\9{{\sst{(9)}}}
\def\p{{\sst{(p)}}}
\def\q{{\sst{(q)}}}
\def\ve{\varepsilon}
\def\vf{\varphi}
\def\F{\Phi}
\def\wg{\wedge}

\def\thb{\bar{\theta}}
\def\Thb{\bar{\Theta}}
\def\barp{\bar{p}}
\def\barq{\bar{q}}
\def\barc{\bar{c}}
\def\bard{\bar{d}}
\def\e{\epsilon}

\def \bi{\bibitem}
\def \la {\label}

\def \l {\lambda}
\def\foot{\footnote}
\def \tl  {{\tilde \l}}
\def \sql {{\sqrt \l}}
\def \adss {$AdS_5 \times S^5$\ }
\newcommand{\rf}[1]{(\ref{#1})}
\def \ov {\over}

\def\th{\theta}
\def\Th{\Theta}
\def\vth{\vartheta}
\def\btheta{{\bar\theta}}
\def\ttheta{{{\tilde\theta}}}
\def\bttheta{{{\bar\ttheta}}}
\def\vth{\vartheta}

\def\ra{\rightarrow}
\def\N{\nabla}
\def\F{{\cal F}}
\def\uM{\underline{M}}
\def\uA{\underline{A}}
\def\uN{\underline{N}}
\def\uP{\underline{P}}
\def\ua{\underline{a}}
\def\ub{\underline{b}}
\def\uc{\underline{c}}
\def\ud{\underline{d}}
\def\ue{\underline{e}}
\def\uf{\underline{f}}
\def\ui{\underline{i}}
\def\uj{\underline{j}}
\def\uk{\underline{k}}
\def\ul{\underline{l}}
\def\ual{\underline{\alpha}}
\def\ube{\underline{\beta}}
\def\um{\underline{m}}
\def\un{\underline{n}}
\def\up{\underline{p}}
\def\uq{\underline{q}}
\def\ur{\underline{r}}
\def\us{\underline{s}}
\def\umu{\underline{\mu}}
\def\unu{\underline{\nu}}
\def\ula{\underline{\l}}
\def\uka{\underline{\k}}
\def\usi{\underline{\s}}
\def\urh{\underline{\r}}
\def\cc{\circ}
\def\eqv{\equiv}

\def\ni{\noindent}

\def\Ep{E^{{}^{(+)}}}
\def\Em{E^{{}^{(-)}}}

\def\Mp{M^{{}^{(+)}}}
\def\Mm{M^{{}^{(-)}}}

\def \ha{{1\ov 2}}

\def\r{\rho}

\def\Y{{\rm Y}}
\def\X{{\rm X}}
\def\tY{\tilde{\rm Y}}
\def\tX{\tilde{\rm X}}
\def\dY{\dot{\rm Y}}
\def\dX{\dot{\rm X}}

\def \J {\mathcal{J}}
\def \del {\partial}

\def\dF{\dot{F}}
\def\dG{\dot{G}}
\def\df{\dot{f}}
\def \E {{\cal E}}
\def \S {{\cal S}}
\def \J {{\cal J}}

\def\ms{\mathcal{S}}
\def\mj{\mathcal{J}}
\def\soj{\fr{\ms}{\mj}}
\def \R {{\bf R}}
\def \om {\omega}
\def \bE {\bar E}
\def \x {{\cal X}}

\def \bi{\bibitem}
\def \la {\label}

\def \l {\lambda}
\def\foot{\footnote}
\def \tl  {{\tilde \l}}
\def \sql {{\sqrt \l}}
\def \adss {$AdS_5 \times S^5$\ }
\def \ov {\over}

\def \varpi {{\rm w}}

\def\thb{\bar{\theta}}
\def\Thb{\bar{\Theta}}
\def\mb{\bar{\m}}
\def\ab{\bar{\a}}
\def\zb{\bar{z}}
\def\psib{\bar{\psi}}
\def\barp{\bar{p}}
\def\barq{\bar{q}}
\def\barc{\bar{c}}
\def\bard{\bar{d}}
\def\e{\epsilon}
\def\wb{\bar{w}}
\def\lb{\bar{\l}}
\def\Jb{\bar{J}}
\def\Nb{\bar{N}}
\def\Zb{\bar{Z}}
\def\pab{\bar{\pa}}

\def\At{\tilde{A}}
\def\Bt{\tilde{B}}
\def\Ct{\tilde{C}}
\def\Dt{\tilde{D}}
\def\Et{\tilde{E}}
\def\Ft{\tilde{F}}
\def\Gt{\tilde{G}}
\def\Ht{\tilde{H}}
\def\Kt{\tilde{K}}
\def\Mt{\tilde{M}}
\def\Nt{\tilde{N}}
\def\Rt{\tilde{R}}
\def\at{\tilde{a}}
\def\bt{\tilde{b}}
\def\ct{\tilde{c}}
\def\dt{\tilde{d}}
\def\et{\tilde{e}}
\def\ft{\tilde{f}}
\def\htil{\tilde{h}}
\def\gt{\tilde{g}}
\def\nt{\tilde{n}}
\def\mut{\tilde{\mu}}
\def\nut{\tilde{\nu}}
\def\pht{\tilde{\f}}

\def\rht{\tilde{\rho}}

\def\asth{\hat{*}}
\def\phh{\hat{\phi}}

\def\bA{{\bf A}}

\def\ola{\overleftarrow}
\def\ora{\overrightarrow}
\def\alt{\tilde{\a}}

\def\eh{\hat{e}}
\def\eph{\hat{\e}}
\def\ph{\hat{p}}
\def\alh{\hat{\a}}
\def\beh{\hat{\b}}
\def\gah{\hat{\g}}
\def\Fh{\hat{F}}
\def\muh{\hat{\m}}
\def\nuh{\hat{\n}}
\def\thh{\hat{\th}}
\def\rhh{\hat{\r}}
\def\dh{\hat{d}}
\def\ih{\hat{i}}
\def\jh{\hat{j}}
\def\hh{\hat{h}}
\def\kh{\hat{k}}
\def\deh{\hat{\d}}
\def\wh{\hat{w}}
\def\lah{\hat{\l}}
\def\Ah{\hat{A}}
\def\Kh{\hat{K}}
\def\Rh{\hat{R}}
\def\Ch{\hat{C}}
\def\Omh{\hat{\Omega}}

\def\xh{\hat{x}}

\def\ps{\rlap{\, /}\;\,p }
\def\ks{\rlap{\, /}\;\,k }

\def\gym{g_{YM}}

\def\adot{\dot{a}}
\def\bdot{\dot{b}}
\def\bpa{\bar{\pa}}

\def\pr{\prime}
\def\ssk{\medskip}
\def\clb{\color{blue}}
\def\clr{\color{red}}
\def\clg{\color{green}}

\begin{document}

\overfullrule=0pt
\parskip=2pt
\parindent=12pt
\headheight=0in \headsep=0in \topmargin=0in
\oddsidemargin=0in

\vspace{ -3cm}
\thispagestyle{empty}

 \vspace{0.1cm}

\setcounter{equation}{0}
\setcounter{footnote}{0}
\setcounter{section}{0}

\begin{center}

{\Large\bf Reduction of BTZ spacetime to hypersurfaces of foliation}

\vskip 0.8cm

 \vspace{.5cm}

\vspace{0.5cm}
I. Y. Park
\\

\vspace{0.1cm}
{\it Department of Applied Mathematics,
Philander Smith College 
                               \\
Little Rock, AR 72223, USA \\
inyongpark05@gmail.com
}

\end{center}

 \vspace{0.1cm}

 \begin{abstract}
We reduce the BTZ spacetime to two kinds of hypersurfaces of foliation: one having a fixed radial coordinate and the other a fixed angular coordinate. The radial reduction leads to a Liouville type theory, and confirms, from the first principle, the expectation laid out in the literature. In the other endeavor, the angular reduction of the 3D gravity is carried out in two different ways; the first again yields a Liouville type theory (different from that of the radial reduction) and the second yields a 2D interacting quantum field theory with quartic potential. We discuss potential implications of our result for the Equivalence Principle and Purity of Hawking radiation.

\end{abstract}
\newpage

\section{Introduction}

A spherically symmetric black hole (BH) is one of the solutions of gravity theory featuring a high degree of symmetry.
It should be possible to describe the dynamics of the branch of the moduli space associated with such a solution by choosing a convenient hypersurface foliation of the bulk configuration. The hypersurface theory is {\em not} always gravity theory, a remarkable feature of gravitational physics well captured by the technique of dimensional reduction to a hypersurface of foliation or ADM reduction for short.

The notion of ADM reduction was first put forth in \cite{Hatefi:2012bp} in the type IIB context, building on the earlier works of \cite{Sato:2002kv,Sato:2003ky,Sato:2004ic}; it was shown  \cite{Sato:2002kv,Sato:2003ky,Sato:2004ic}\cite{Hatefi:2012bp}  that the hypersurface theory is an abelian gauge theory. Since the hypersurface dynamics should capture aspects of the bulk physics, the hypersurface theory can be viewed as a ``dual" description of the bulk physics in the sense discussed in \cite{Hatefi:2012bp}. If this view is correct (as we believe), it would lead to
plethora of dual pairs of gravity/non-gravity theories. This generalized duality, when more firmly established, should provided a first-principle derivation of one of the two facets of AdS/CFT: how the open string physics emerges from the closed string physics. The idea is also consistent with the view that this facet of AdS/CFT should be a generalized spontaneous symmetry breaking phenomenon \cite{Park:2008sg}.

Dimensional reduction to various hypersurfaces has proven fruitful. In the big picture, it provides
a unifying paradigm for generating holographic dual pairs. In the initial work of \cite{Hatefi:2012bp}, the notion was conceived in the IIB supergravity setup by building on \cite{Sato:2002kv}, in which
a DBI type action was obtained as a solution of IIB supergravity through Hamilton-Jacobi formalism. The gauge field was interpreted as a ``moduli field"
that describes the fluctuations of the branch of the supergravity moduli space associated with the $S^5$ reduction. The DBI action should capture dynamics of the supergravity and, naturally, can be viewed as the dual theory. It is almost evident that the non-abelianized DBI action would be the stringy extension of ${\cal N}=4$ $D=4$ SYM. (The non-abelianization procedure was sketched in \cite{Hatefi:2012bp}; although it would require a substantial amount of work, it will be doable.)
Once the non-abelian extension is completed, the
procedure will be a derivation (possibly up to some fine issues) of AdS/CFT\footnote{The theory dual to IIB closed string
will be open stringy extension of ${\cal N}=4$ $D=4$ SYM in the curved background \cite{Park:1999xz}.} to the extent that the moduli field interpretation is valid. (There is no doubt that the moduli field interpretation carries a certain level of truth.)

One of the virtues of the ADM reduction is that it is not limited to stringy setups, but much more general.
The ADM reduction has been applied to the 4D Schwarzschild black hole in \cite{Park:2013iqa} and \cite{Park:2013vpa}.
 The Schwarzschild spacetime was reduced along the radial direction in \cite{Park:2013iqa} and
along the angular directions in \cite{Park:2013vpa}. Once the 4D theory is reduced to the hypersurface located at the event horizon, the theory effectively becomes a 2D theory and the entropy (or the zero temperature analogue of entropy) can be computed. The angular reduction was considered in order to obtain a setup optimized for analyzing scattering around the black hole at the level of {\em interacting quantum field theory} (QFT) that describes gravity physics in the hypersurfce; namely, the ``quantum gravity of the hypersurface".

In this work, we apply the procedure to another case, the BTZ black hole \cite{Banados:1992wn}.
The ADM reduction and subsequent analysis of the BTZ spacetime are expected
to be simpler than those of the Schwarzschild spacetime for several reasons. Firstly, it obviously takes reduction along a single direction to reach a 2D theory. Secondly, conformal symmetry
has been expected to emerge in the 2D theory, and if it indeed does, conformal field techniques can be employed for various further computations. The conformal field theory expected is Liouville theory in the radial reduction case.
The connection between the BTZ solution and Liouville theory was discussed, e.g., in \cite{Coussaert:1995zp,Frolov:1999my,Krasnov:2000ia,Giacomini:2003cg,Chen:2003si,
Yuan:2011gq,Nakatsu:1999wt}. Appearance of Liouville theory was shown in \cite{Solodukhin:1998tc} (see \cite{Cvitan:2002cs,Giacomini:2004ts} as well) in the context of the standard dimensional reduction of 4D actions.

Below, these expectations will be confirmed. We show the emergence of a Liouville type theory in the radial reduction; as a byproduct we will be able to clarify some of the less-understood issues. (For example, whether the Liouville theory is associated with the horizon or the asymptotic region is not yet completely understood in the literature \cite{Carlip:1998qw}.) The angular reduction also leads to a Liouville type theory that is different from that of the radial reduction.

\begin{figure}
\centerline{
\begin{minipage}[b]{9cm}
             \epsfxsize=9cm
              \epsfbox{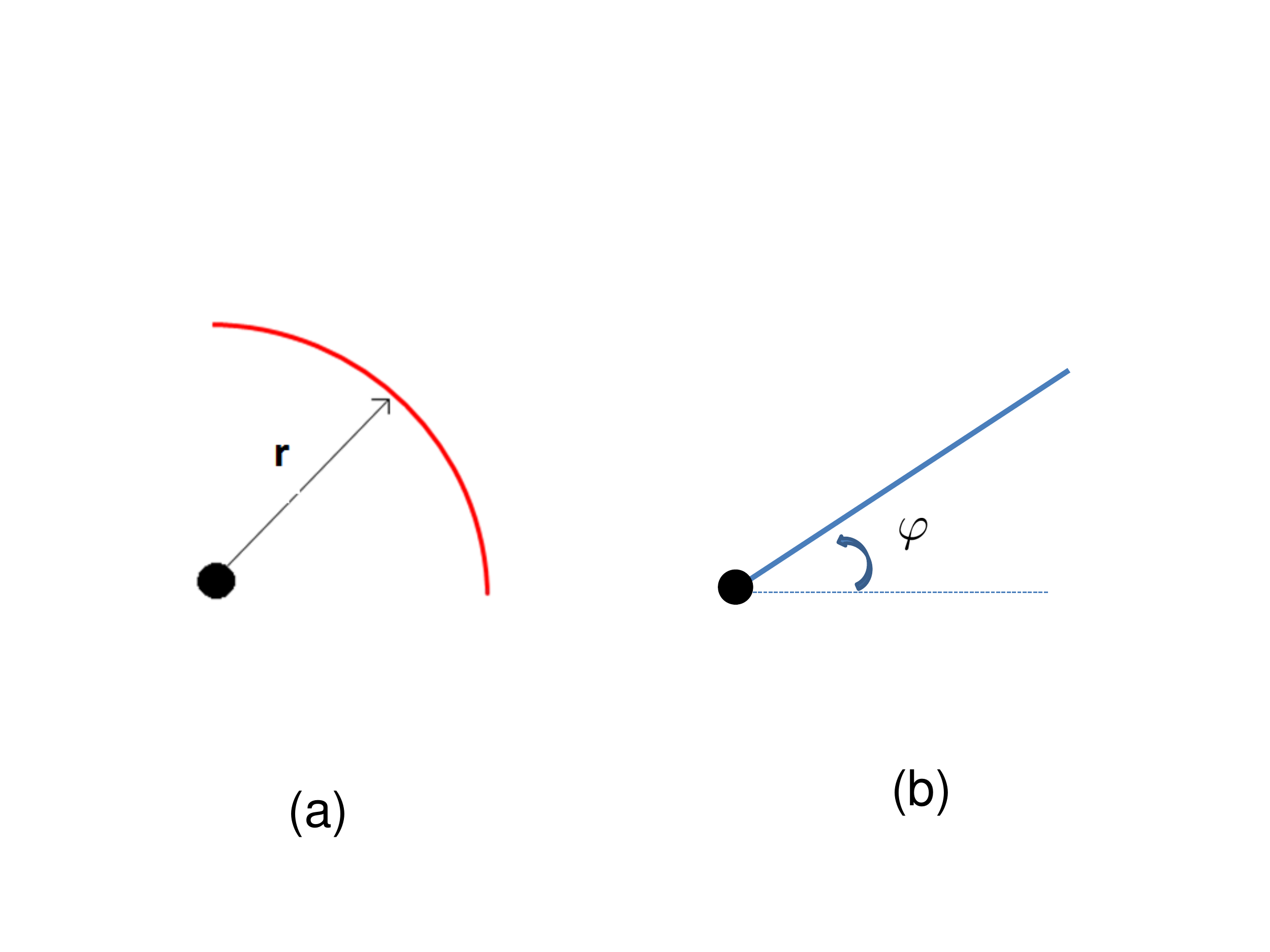}
      \end{minipage}
      }
\caption{(a) $r$ reduction  (b) $\vf$ reduction}
\label{fig}
\end{figure}

The resulting 2D theories are interacting QFTs up to the issue that we will discuss in the last section. They describe the metric fluctuations in or around the hypersurfaces, i.e., the quantum gravity theories (once quantized) of the hypersurfaces.
With the second-quantizable theories at hand, we examine notions/pictures that are
either semi-classical or have roots in the semi-classical description from the perspective of the present work.

\vspace{.3in}
The remainder of the paper is organized as follows.
In section 2, we carry out the radial reduction.\footnote{
The hypersurface analysis that we carry out here (and have carried out in \cite{Park:2013iqa} and \cite{Park:2013vpa}) does not describe the full bulk dynamics. To elucidate the point, let us take the radial reduction. The reduced theory does not describe the collective motions of the surfaces foliated along the radial direction. Perhaps an analogy would help. In the atomic theory of a hydrogen atom, one factors out the motion of the center of mass, and focuses on the oscillations of the electron and proton around the center of mass. The part that is not described by the reduced theory should be analogous to the motion of the center of mass. The full bulk dynamics would definitely include this part.
} and obtain a Liouville type theory.
The reduction along the azimuthal angle is carried out in two different approaches in section 3. The first approach yields a Liouville type theory. In the second approach, the resulting 2D theory has quartic potential in terms of a redefined field. In the last section, we discuss potential implications of our result for
the Equivalence Principle (EP) and Purity of Hawking radiation.

\section{$r$-reduction and Liouville theory}

Let us consider the 3D Einstein-Hilbert action with a cosmological constant
\be
S=\int\, d^3x\, \sqrt{-\gt^\3}\left[\Rt^\3-2\L \right]
\la{AdS3act}
\ee
where
\be
\L=-1/l^2
\ee
with $l$ being the AdS characteristic length.
The field equation admits the following black hole solution \cite{Banados:1992wn}
\be
ds^2=-f(r)dt^2+f(r)^{-1}dr^2+r^2 d\vf^2,\qquad f(r)\equiv \fr1{l^2}(r^2-r^2_H)
\la{3DBH}
\ee
where $r_H$ denotes the radial location of the event horizon.
Let us employ the ADM formalism\footnote{Below we will go back and forth between the standard formulation and ADM formulation as need arises. }
\bea
ds_3^2=(\nt_r^2+\htil^{\2 ij}\Nt_i \Nt_j)dr^2+2\Nt_i dr dy^i+\htil_{\2 ij}dy^idy^j
\eea
where $\htil_{\2 ij}$ is the metric of the hypersurface and $\Kt_{ij}\equiv \fr{1}{{2\nt_r}}(\pa_r \htil_{\2 ij}-\tilde{\N}_i \Nt_j-\tilde{\N}_j \Nt_i)$, $\Kt\equiv \htil_\2^{ij}\Kt_{ij}$. $\Nt_i$ can be gauge-fixed away\footnote{
The gauge-fixing should amount to narrowing down to the branch of the moduli space in which the shift vector is absent.
};
the 3D action can be written
\bea
S&=&\int  d^2y  dr \, \sqrt{-\htil_\3}\left[\Rt^\2+\Kt^2-\Kt_{ij}\Kt^{ij}
 -2\L\right] \la{3drescor}
\eea
It is convenient to redefine
\bea
\nt_r(t,r,{ \vf})\equiv e^{{ -}\pht(t,r,{ \vf})}
\eea
As an intermediate step toward the dimensional reduction to 2D (in other words, \rf{phired} will be part of the reduction ansatz below), let us narrow down to a sector of the theory where $\nt_r$ takes
$\nt_r$ takes
\bea
e^{\pht(t,r,{ \vf})}= e^{\f_0(r)+\f(t,{ \vf})}= { \sqrt{f}}e^{\f(t,{ \vf})} \la{phired}
\eea
where $e^{{\f_0}}$  has been set { $e^{{\f_0}}={\sqrt{f}}$} in order to assure the fluctuations around the BTZ spacetime \rf{3DBH}.
 Separating out the ``breathing mode", rescale the 3D metric $\htil_{\m\n}$
  by\footnote{This is a non-covariant expression in 3D since $e^{\f}$ is
 not a scalar. However, this should not cause a probelm once the theory is reduced to 2D where $e^\f$ becomes a 2D scalar.}
\bea
\htil_{\m\n}=e^{2{ \f(t,\vf)}}h_{\m\n}   \la{hres}
\eea
 The action becomes
\bea
S&=&\int  { d^2y}dr \, \sqrt{-h_\2}\;{ \fr{1}{\sqrt{f}}}
\left[R^\2+K^2-K_{ij}K^{ij}
-4\N_\3^2\f-2(\N_\3\f)^2
 +\fr{2}{l^2} e^{{ 2}\f}\right] \nn\\  \la{3drescor3}
\eea
The action may be modified by surface terms
in order to accommodate virtual boundary effects \cite{Park:2013vpa}. (The presence of the usual Gibbons-Hawking type boundary terms are also understood.) We will carry out this analysis shortly but first let us proceed without worrying about the boundary terms.
One can show that the 3D field equations are reduced to
a set of equation that can be produced from
\bea
S&=&\int  { d^2y} \, \sqrt{-h_\2}\;{ \fr{1}{\sqrt{f}}}
\left[R^\2   
-2(\N_\2\f)^2
 +\a\f
 +\fr{2}{l^2} e^{{ 2}\f}\right]
 \nn\\\la{3drescor4}
\eea
  Here $K_{ij}\equiv \fr{1}{{2}}e^{{ 2\f}}{ \pa_r h_{ij}}$, $K\equiv h^{ij}K_{ij}$ and
\bea
\a\equiv -4\sqrt{f}\;\N_\3^2\fr{1}{\sqrt{f}}
   +4\sqrt{f}\N^\m\Big(\fr{1}{\sqrt{f}}\N_\m \f_0\Big)
\eea
This can be viewed as a constant since $r$ is no longer a coordinate but a parameter.
 To derive \rf{3drescor4}, we have used $\Rt^\2+(\Kt^2-\Kt_{ij}\Kt^{ij})=\Rt^\3$ (up to total derivative terms) and how $\Rt^\3$ rescales under \rf{hres}.
The $K^2-K_{ij}K^{ij}$ term vanishes after the 2D metric gauge fixing as we show in \rf{Ksvan} below, and has been omitted.

As mentioned above, the action \rf{3drescor4} is valid up to the terms that originate from the total derivative terms added to \rf{3drescor3}. The fact that not all is well
can be seen as follows. With the metric rescaling \rf{hres},  the system is expected admit the following solution
\bea
 \f(t,\vf)=0
\eea
\bea
  h_{ij}=
\left(
\begin{array}{cc}
 -1  & 0  \\
 0 & \fr{r^2}{f(r)}
\end{array}
\right) (\equiv \g_{0ij} \la{fgsol})
\eea
One can easily see that the naive form of the action \rf{3drescor4} does not admit \rf{fgsol} as a solution. For example, one can quickly see that the $\f$ equation does not admit $\f=0$. As a matter of fact, the action is defined up to
shifting $\f$ by a constant. This is because the change in the
cosmological constant term causd by the shift in $\f$ can be absorbed by
another source for the cosmological term as we turn now.

 Examination of \rf{3drescor3} reveals the forms of the needed boundary term, and
 its precise coefficient is determined by requiring \rf{fgsol} as a solution.
Suppose taking $h_{ij}^\3$ variation of the action after going back to the standard formulation: $R^\2+(K^2-K_{ij}K^{ij})=R^\3$. The resulting field equation contains
a term $\sim \N_\3^2 \fr{1}{\sqrt{f}}$. (Note that $\N_\3^2 \fr1{\sqrt{f}}=\fr{2r_H^2}{l^4}\fr1{\sqrt{f}}$.) In other words, the action \rf{3drescor4}
has been determined up to the boundary effect that is represented by the presence of $\sim \N_\3^2 \fr{1}{\sqrt{f}}$ in \rf{3drescor3}.

With all of the observations above taken into account, the 2D action is given by
\bea
S&=&\int  { d^2y} \, \sqrt{-h_\2}
\left[R^\2  -2(\N_\2\f)^2+\a\f +\fr{2}{l^2} e^{{ 2}\f}+\k\right]\la{3drescor6}
\eea
where $\k$ is a constant (or a $r$-dependent expression, more precisely speaking).
The metric still has the $r$-dependence; however, $r$ now serves as a parameter rather than a coordinate as mentioned above. For the description of 2D dynamics, let us gauge-fix the 2D metric (the Virasoro constraint should be supplemented to the action)
\bea
 ds^2_2 = \g_{0ij}(r) dy^i dy^j,\;\; (\mbox{i.e.,}\;\; h_{ij}=\g_{0ij})
\la{adsansr2q}
\eea
where $\g_{0ij}$ was defined in \rf{fgsol}.
Note that in this gauge
\bea
R^\2=0,\quad (k^2-k_{ij}k^{ij})=0  \la{Ksvan}
\eea
where $k_{ij}\equiv \fr{1}{{2}}{ \pa_r h_{ij}}$, $k\equiv h^{ij}k_{ij}$. (As a matter of fact, $kk_{ij}-k_{im}\g^{ml}k_{lj}=0$.)
Removing the constant terms followed by rescaling, one gets
\bea
S&=&\int  dtd\vf  \left[
   -\fr12\g_0^{ij}(\pa_i\f)(\pa_j\f) 
   +\b \f
   +\fr1{2l^2} e^{2\f}\right] \la{3drescor3red2}
\eea
where $\b$ is rescaled $\a$.
 Once the hypersurface is set at the event horizon, the theory will effectively be reduced by one more dimension as in \cite{Park:2013iqa}. Because of this reason, the Liouville theory in the context of
BTZ spacetime should presumably not be associated with the horizon.
One can rescale the coordinates $(t,\vf)$ and the field $\f$, and put the action into a more
standard-looking form.

\section{$\vf$-reduction: preliminary scattering-setup}

As discussed in \cite{Park:2013vpa}, reduction along the angular direction may provide a convenient setup for analysing scattering around the black hole.
We carry out the reduction in two different ways. In the first approach, we closely follow the steps of the $r$-reduction, and obtain another Liouville type theory. In the second approach, which might provide a more convenient setup for scattering analysis,
we obtain a theory that has a tachyonic mass and quartic potential in terms of an appropriately redefined field.
 The kinetic terms of these theories have $r$-dependent coefficients; this feature, with the Virasoro type constraint, will complicate the scattering amplitude analysis. Nevertheless, the 2D actions takes relatively simple forms, and
 quantization and scattering analysis (to be pursued elsewhere) should be reasonably straightforward. Since the terms that originate from the virtual effects do not usually play an important role as we saw in \cite{Park:2013vpa} and in the previous section, we do not keep track of them in this section.

\subsection{$\vf$-Reduction: approach 1}

By employing the ADM formalism
\bea
ds_3^2=(\nt_\vf^2+\htil^{ab}\Nt_a \Nt_b)d\vf^2+2\Nt_a d\vf dx^a+\htil_{ab}dx^adx^b
\eea
the 3D action can be written
\bea
S&=&\int  d^2x  d\vf \, \sqrt{-\htil}\;\nt_\vf\left[\Rt^\2+\Kt^2-\Kt_{ab}\Kt^{ab}
 -2\L\right] \la{3drescorphi}
\eea
$\Nt_a$ is {gauge-fixed} to $\Nt_a=0$.
Redefine the field ${\tilde{n}}_\vf(t,r,{ \vf}) \equiv e^{{ -}\rht(t,r,{ \vf})}$.
Rescale the 3D metric $\htil_{\m\n}$ by
\bea
\htil_{\m\n}=e^{2\rht(t,r,\vf)}h_{\m\n}
\eea
One gets, after
reduction to 2D by setting $\rht(t,{ r},{ \vf}) =\r_0(r)+  \r(t,r)$ with $e^{-\r_0(r)}=r$, 
\bea
S&=&\int  d^2x    \sqrt{-h}\;\left[R^\2+\a_\vf(r) \r(t,r)
   -2(\N_a\r)^2+(K^2-K_{ab}K^{ab})
 +\fr2{l^2} {e^{2\r_0}}e^{2\r(t,r)}\right] \nn\\\la{3drescor3q}
\eea
where $\a_\vf(r)$ is a $r$-dependent expression that is determined up to the freedom
of shifting $\r$.
Let us gauge-fix the 2D metric (the Virasoro constraint should be supplemented as usual)
\bea
ds^2_2=\g_{0ab} dx^a dx^b
\la{adsansr2q2}
\eea
where
\bea
\g_{0ab}\equiv
\left(
\begin{array}{cc}
 -r^2f(r) & 0  \\
 0 & \fr{r^2}{f(r)} \\
\end{array} \la{gamz2}
\right)
\eea
After going through the steps analogous to the ones in the previous section, the action takes
\bea
S&=&\int  d^2x    \sqrt{-\g_0}\left[
   -2(\N_a\r)^2+\a_\vf(r) \r(t,r)
 +\fr{2}{l^2}{e^{2\r_0}}e^{2\r(t,r)}-\fr{2}{l^2}\right] \la{3drescor3q2}
\eea
which can be rewritten
\bea
S&=&\int  d^2x   \sqrt{-\g_0} \left[
   -\fr12\g_0^{ab}\pa_a\r \pa_b\r +\fr14\a_\vf(r) \r(t,r)
 +\fr1{2l^2} {e^{2\r_0}}e^{2\r(t,r)}\right] \la{3drescor3q2}
\eea
where the action has been numerically rescaled and the field independent terms have been removed.

\subsection{$\vf$-Reduction: approach 2}

In this subsection, we carry out a slightly different $\vf$ reduction.
 This approach leads to a theory with a quartic potential that may provide a more convenient setup for the scattering analysis. (It also allows one to see the free field theory form of the resulting action, a point
 that we will discuss in the discussion section.\footnote{As we will see, there exists a different rescaling that casts \rf{vfredactsim} into a free theory.
Possible implications will be discussed in the final section. Interestingly, it was shown in \cite{Thorn:1983qr}\cite{D'Hoker:1983is}\cite{Yoneya:1984dd} that Liouville
theory can be mapped to a free field theory.   \la{lff}
})

With the following block-diagonal gauge-fixing
\bea
ds^2=\nt_{\vf}d\vf^2+\htil_{ab}dx^adx^b, \quad a=(t,r)
\la{metgfvf}
\eea
the 3D action can be written
\bea
S &=& \int   d^2xd\vf \, \;\nt_{\vf}\sqrt{-\htil}\Big[\Rt^\2+\Kt^2-\Kt_{ab}\Kt^{ab}  -2\L \Big]
\la{adsrredresvf}
\eea
Redefining the field $\nt_\vf$ by
\bea
\nt_\vf(t,r,{ \vf}) \equiv e^{\r(t,r,{ \vf})}
\eea
and reducing the 3D action \rf{adsrredresvf} to 2D by
\bea
\r(t,{  r},{ \vf}) &\ra & \r(t,r)
\eea
one gets
\bea
S &=& \int   d^2x\, e^\r\sqrt{-\htil}\Big[\Rt^\2+\Kt^2-\Kt_{ab}\Kt^{ab}  -2\L \Big]
\la{adsrredresvfred}
\eea
As before, the breathing mode can be separated out by the rescaling
\bea
\htil_{ab}=e^{\r(t,r)}h_{ab} \la{habres1}
\eea
and the action becomes
\bea
S=\int \,  dtdr \sqrt{-h}\, e^\r\left[R^\2+K^2-K_{ab}K^{ab}
- \N_a\N^a \r  +\fr2{l^2}  e^{\r} \right] \la{freeact}
\eea
Let us gauge-fix
\bea
ds^2_2=\g_{0ab} dx^a dx^b
\la{adsansr2q}
\eea
where the 2D background metric is given by
\bea
\g_{0ab}\equiv
\left(
\begin{array}{cc}
 -f(r) & 0  \\
 0 & \fr{1}{f(r)} \\
\end{array} \la{gamz}
\right)
\eea
The action simplifies to
\bea
S=\int \,  dtdr \sqrt{-\g_0}\,\left[e^\r (\N_a \r)^2 -\fr2{l^2} e^\r
 +\fr2{l^2}  e^{2\r} \right] \la{rhoact}
\eea
where the fact $R^\2=-\fr2{l^2}$ has been used. The corresponding Virasoro constraint should be supplemented.
Noting that $\sqrt{\g_0}=1$, the action \rf{rhoact} takes a slightly simpler form
\bea
S=\int   dt \int_{r_H}^\infty dr\; \left[e^\r (\N_a \r)^2 -\fr2{l^2} e^\r
 +\fr2{l^2}  e^{2\r}\right]
\la{vfredactsim}
\eea
with the Virasoro constraint
\bea
\pa_a\r\pa_b\r-\fr12 \g_{0ab}\Big[(\pa \r)^2 +\fr2{l^2}(e^\r-1) \Big]=0
\eea

To see that the action above can be written as a theory with a quartic potential,
let us introduce a further field redefinition
\bea
e^{\fr12 \r}\equiv \xi
\eea
which casts the action to
\bea
S=\int   dt \int_{r_H}^\infty dr\; \left[4 (\pa_a \xi)^2 -\fr2{l^2} \xi^2
 +\fr2{l^2}  \xi^4\right]
\la{actinxi}
\eea
The Virasoro constraint takes
\bea
\pa_a \xi\, \pa_b \xi- \fr12\g_{0ab}\Big[(\pa \xi)^2+\fr1{2l^2}(\xi^4-\xi^2) \Big]=0
\la{vira1}
\eea
The $\xi$ field equation is
\bea
\pa^2 \xi+\fr{1}{l^2}(\fr12\xi-\xi^3)=0 \la{xieom}
\eea
Upon using this in \rf{vira1}, the constraint becomes
\bea
\pa_a \xi\, \pa_b \xi- \fr12\g_{0ab}\Big[(\pa \xi)^2
    +\fr1{2} {\xi \pa^2 \xi}-\fr1{4l^2}\xi^2 \Big]=0
\eea
It should be useful to consider a time-dependent solution of \rf{xieom}.
\ssk\ssk

Prior to scattering analysis, one must understand whether \rf{vfredactsim} represents a truly interacting theory because there exists a field redefinition that
casts \rf{freeact} to a free field form.
We take this discussion in the next section.

\section{Discussion}

In this work, we have carried out reduction of BTZ spacetime to two different kinds of hypersurfaces, one with the fixed radial coordinate and the other with the fixed angular coordinate. The radial reduction has led to \rf{3drescor3red2} and the angular reduction to \rf{3drescor3q2} and \rf{actinxi}. The resulting 2D theories provide a basis for a second-quantized description -which should be essential for tackling
puzzles such as Black Hole Information paradox and Firewall \cite{Almheiri:2012rt} - of fluctuations around the black hole in the hypersurface.

With the second quantizable actions available, we are now in a position to carry out direct quantum field theory analysis, and we will report on the progress elsewhere. In the remainder of this section, we ponder over various issues that will emerge when setting a strategy for relating scattering analysis to the Black Hole Information and Firewall.

The second quantized description of gravitational fluctuations
would require, as suggested by \cite{Almheiri:2012rt}, one to re-scrutinize several key notions such as the Equivalence Principle and Purity of Hawking radiation. This is because they are notions deeply rooted in the semi-classical description, which treats only the matter fields in the quantum manner with the metric serving as a fixed background. Furthermore, mostly free matter field theories were employed in the BH literature of this context.

We start with Equivalence Principle.
As stated in the main body, there exists a frame that leads to free field theory in the angular reduction case. To see this, let us rescale the metric in \rf{freeact} by
\bea
h_{ab}= e^{- \r}\g_{ab} \la{fres}
\eea
Upon this rescaling, \rf{freeact} becomes
\bea
S=\int \,  dtdr \sqrt{-\g}\,  e^\r\left[R^\2+K^2-K_{ab}K^{ab}
- \N_a\N^a \r  +\fr2{l^2}  \right]
\eea
Once the metric is gauge-fixed, the action becomes
\bea
S &=& \int \,  dtdr \sqrt{-\g_0}\; e^\r \Big(- \N_a\N^a \r  \Big)
   = \int \,  dtdr \sqrt{-\g_0}\; 4  (\N \z)^2
\la{vfredactfree}
\eea
where $\z\equiv e^{\fr12 \r}$.
The field redefinition \rf{fres} would amount to conducting a very special coordinate transformation in the original 3D theory. It implies that there is a certain frame
 in which the QFT interactions are removed.
The redefined coordinate would correspond to an observer in a novel frame.
Whether this feature is a general phenomenon is far from obvious, of course.
(It is interesting, though, that Liouville theory was shown to map to a free field theory as mentioned in footnote \rf{lff}.)
If it is, it might be possible to interpret the removal of the gravitational interactions (recall that the QFT is theory of a {\em gravity mode} in the hypersurface) by a coordinate transformation as EP of a certain kind.\footnote{ Even if it is indeed a modified EP, there is a question whether it has any useful contents. This is because the special frame does not seem to be related to the Kruskal coordinate.
}
However, the theory is free when expanded around a {\em novel} vacuum and the vacuum does {\em not} seem to be the Kruskal vacuum: the theory must be an interacting theory when expanded around the Schwarzschild or Kruskal vacuum.

The interacting quantum field theory nature of the fluctuating geometry also
seems to suggest the possibility of information bleaching - which was the central motivation of the work \cite{Park:2013rm} - at the black hole formation and/or growth.
An infalling chair might go through a ``gravitational Bremsstrahlung" process by which its main (or at least partial) pieces of information might get bleached to the vicinity of the event horizon.
If indeed present, the information bleaching mechanism will dramatically change (the order of) the questions regarding the information: before exploring whether or how information escapes, one would examine what pieces of information actually enter the black hole. Information bleaching would be at odds with Purity
of Hawking radiation because there would not be much information to carry to the outside.\footnote{But the combined system of the Hawking radiation, ``non-Hawking radiation" and jets may be pure \cite{Park:2013rm}.
}
Another semi-classical picture that might not be fully supported by the quantum gravity description is the process of disintegration of a fallen chair throughout which the BH remains {\em entirely} black. As far as we can see, there is a possibility that
the black hole goes through a series of ``meta-black" quantum states between the initial and final black states, a hypothetical process called ``blackening" in \cite{Park:2013rm}. We will report the progress on these issues in the near future.

\vspace{.1in}

\ni {\bf Acknowledgements}

\ni The author thanks A. Nurmagambetov and J. Polchinski for useful discussions.

\newpage

\end{document}